\documentclass[twocolumn,showpacs,preprintnumbers,amsmath,amssymb,superscriptaddress]{revtex4}

\usepackage{graphicx}
\usepackage{dcolumn}
\usepackage{bm}
\usepackage{amsmath}

\begin{document}

\preprint{LA-UR-05-3141}

\title{Do solar neutrinos constrain the electromagnetic properties of
  the neutrino?}

\author{Alexander Friedland} \email{friedland@lanl.gov}
\affiliation{%
  Theoretical Division, T-8, MS B285, Los Alamos National Laboratory,
  Los Alamos, NM 87545
}%

\date{September 7, 2005}

\begin{abstract}
  It is of great interest whether the recent KamLAND bound on the flux
  of electron antineutrinos from the Sun constrains the
  electromagnetic properties of the neutrino. We examine the
  efficiency of the electron antineutrino production in the solar
  magnetic fields, assuming the neutrinos are Majorana particles with
  a relatively large transition moment. We consider fields both in
  the radiative and convective zones of the Sun, with physically
  plausible strengths, and take into account the recently established
  values of the oscillation parameters. Our analysis shows that the
  production rate in question is presently unobservable. In the
  radiative zone, it is suppressed by the large measured value of the
  flavor mixing angle which eliminates the resonant level crossing. A
  corresponding general resonance condition, valid for large as well
  as small values of the mixing angle, is derived.  Likewise, in the
  convective zone, the strength of the small-scale magnetic field is
  likely to be insufficient.  Thus, no useful bound on the neutrino
  transition moment can be derived from the published KamLAND bound.
  KamLAND may be, however, on the edge of probing an ``optimistic''
  scenario, making further improvements of its sensitivity
  desirable.
\end{abstract}

\maketitle

\section{Introduction}

One of the most important current tasks in neutrino physics is to
understand the full implications of the solar
\cite{SK1258,SNOCC,SNONC,SNODN,SKnu2002,SKnubar,SNOsalt,KamLANDnubar,SNOnubar,SNO2005},
atmospheric \cite{SKatm2001,SKatmdip}, and reactor
\cite{KamLANDflux,KamLANDspectrum} neutrino data that became
available in the last five years. While much of the recent effort has been
focused on establishing the values of the neutrino masses and
mixings, there has also been a growing realization that the same data
may contain valuable information about other neutrino properties, such
as its interactions with matter
\cite{Maltoni2001,Guzzo2001,oursolar,brazilians,ConchaMichele2004,valencians,ouratm,inprep}
or with electromagnetic fields
\cite{BeacomVogel1999,AkhmedovPulido2002,ourmagnfit,TorrenteLujan2003,Chauhanetalcore2003,rashbaprl,rashbaprd}.
Any measured deviation of these properties from their Standard Model (SM)
values would have profound theoretical implications.

As part of this program, we wish to reexamine here the sensitivity
of the recent solar neutrino measurements to the neutrino
electromagnetic properties.

The idea that the neutrino may possess a relatively large magnetic
(transition) moment
has been the subject of intense theoretical speculations
for many years. It was originally put forth in the early 1970's, as a
possible explanation to the solar neutrino deficit \cite{Cisneros}:
the solar neutrinos, it was argued, were ``disappearing'' as a result
of the neutrino spin precession in the solar magnetic field.
Remarkably, this explanation -- much improved with time
\cite{Voloshin:1986ty,Okun1986short,Okun1986,Akhmedov1988,LimMarciano,Raghavan}
-- remained viable for the next three decades\footnote{While, by the
  late 1990's, the lack of time variations in the Super-Kamiokande
  data gave strong evidence against spin precession in the solar
  convective zone, spin precession in the radiative zone continued to
  give a good fit to all solar data \cite{ourmagnfit}.}.  Only in 2002
did the KamLAND measurement of the reactor antineutrino disappearance
\cite{KamLANDflux} conclusively rule it out
as the \emph{dominant} mechanism of the solar neutrino conversion.
The possibility of the spin precession happening at a
\emph{subdominant} level, however, remains of great interest to this
day, as a probe of the neutrino electromagnetic properties and, at the
same time, of the magnetic fields in the solar interior.

For definiteness, here we will focus on the case of Majorana
neutrinos.  The experimental sensitivity to magnetic spin precession
in this scenario is especially high, as has been long recognized
\cite{BalantekinLoreti1992}: upon the spin flip, a Majorana neutrino turns
into an antineutrino and -- thanks to flavor oscillations -- can then
be detected with some probability as an electron antineutrino,
$\bar\nu_e$. The $\bar\nu_e$ flux from the Sun, in turn, is strongly
constrained by Super-Kamiokande \cite{SKnubar}, SNO \cite{SNOnubar},
and especially KamLAND \cite{KamLANDnubar} which can detect
$\nu_e\rightarrow\bar\nu_e$ conversion of the solar $^8$B neutrinos at
the level of a few hundredths of a percent.

To understand if the KamLAND bound places a nontrivial constraint on
the neutrino magnetic moment, we need to compare it to the maximum
possible $\bar\nu_e$ flux from the Sun. This must be done taking into
account (i) the recently measured values of the oscillation
parameters, (ii) the bounds on the solar magnetic field strength and (iii)
the existing direct bounds on the neutrino magnetic moment.  We also
need to consider that magnetic spin flip could occur either in the
convective zone (CZ) of the Sun, which comprises the outer 30\% by
radius, or deeper in the radiative zone (RZ). In the first case, the
neutrino traverses a region of turbulent magnetic field; in the second
case, the field, if present, is smooth and frozen into stationary
plasma. The physics of the neutrino evolution is quite different in
the two cases and requires separate treatments.

Both the CZ \cite{TorrenteLujan2003,rashbaprl,rashbaprd} and the RZ
\cite{AkhmedovPulido2002,Chauhanetalcore2003,BalantekinVolpe} cases
have been investigated in the literature and in both cases it has been
concluded that it was {\it a priori} possible to have a signal greater
than the current KamLAND sensitivity reach.
In the case of the CZ, even a bound on
the neutrino magnetic moment was claimed.

In our analysis, we reach different conclusions. In the case of the
RZ, we find that the usual treatment, based on the
$\nu_e\rightarrow\bar\nu_\mu$ resonance, works only for small flavor
mixing. The large measured value of the solar neutrino mixing angle
leads to a qualitatively new effect: the resonant level crossing
disappears and no measurable magnetic spin flip can take place. A new
analytical result, extending the classical ``magnetic resonance''
condition \cite{Akhmedov1988,LimMarciano} to large flavor mixing, is
derived here for the first time.

In the case of the CZ, we consider two qualitatively different models
of the magnetic field. One of these models, that of the uniform
Kolmogorov turbulence, follows \cite{rashbaprl,rashbaprd} and our
basic treatment of the neutrino evolution in this case agrees with
that work.  We, however, take what we regard as more physically
justifiable magnetic field parameters.  For the second model, we take
a scenario of isolated flux ropes as suggested by the recent
solar physics research. This scenario, to the best of our knowledge,
is considered in the neutrino literature for the first time.  The net
result is that in both cases we estimate the maximum $\bar\nu_e$ flux
to be \emph{below} the present KamLAND bound.

We begin by giving the necessary background information, including the
form of the interaction (Sect.~\ref{sect:interaction}), the known
bounds on the neutrino magnetic moment
(Sect.~\ref{sect:directbounds}), the general properties of the solar
magnetic fields (Sect.~\ref{sect:magnfields}), and the form of the
neutrino oscillation Hamiltonian (Sect.~\ref{sect:oscHamiltonian}).
The case of the RZ magnetic field is treated in Sect.~\ref{sect:rz}:
after reviewing the conventional approach
(Sect.~\ref{sect:rz_conventional}), we argue that resonant
antineutrino production is suppressed for large flavor mixing
(Sect.~\ref{sect:rz_nores}), derive a corresponding general analytical
condition (Sect.~\ref{sect:rz_analyt}), and finally perform a direct
numerical scan of the parameter space to confirm our analytical
conclusions (Sect.~\ref{sect:rz_scan}). This is followed by the
analysis of the CZ fields in Sect.~\ref{sect:cz}, where the two models
of the magnetic field mentioned above are considered (in Sects.
\ref{sect:cz_Kolmogorov} and \ref{sect:cz_tubes}). The results are
summarized in Sect.~\ref{sect:conclusions}. Some additional
technical details about the CZ neutrino evolution can be found in the
Appendix.

\section{Background}

\subsection{Interaction}\label{sect:interaction}

The electromagnetic moment interaction of the Majorana neutrinos is
given by the following dimension five operator
\begin{eqnarray}
  \label{eq:lagr4}
  {\cal L}_{EM} &=& -\frac{1}{2}\mu_{a b} \bar\Psi_a
  \Sigma^{\mu\nu}
  \Psi_b F_{\mu\nu},
\end{eqnarray}
where $\Psi_a$ and $\Psi_b$ are Majorana spinors of flavors $a$ and
$b$ in the 4-component notation, and $\Sigma^{\mu\nu}\equiv
i[\gamma^\mu,\gamma^\nu]/2$, with $\mu$ and $\nu$ being the Lorentz
indices.

It proves useful to rewrite this Lagrangian in the 2-component Weyl
spinor notation.  This notation, besides being simply more natural for
ultrarelativistic fermions, brings considerable
physical insight.

Writing the 4-component Majorana spinor $\Psi$ in
the Weyl basis as $\Psi=(\chi_\alpha,\bar\chi^{\dot\alpha})^T$ and
introducing $2\times 2$ matrices $(\sigma^{\mu\nu})_{\alpha}^{\phantom{\alpha}\beta}\equiv i(\sigma_{\alpha\dot\alpha}^{\mu}\bar\sigma^{\dot\alpha\beta\:\nu}-\sigma_{\alpha\dot\alpha}^{\nu}\bar\sigma^{\dot\alpha\beta\:\mu})/2$,
with $\sigma_{\alpha\dot\alpha}^{\mu}\equiv (1,\vec{\sigma})$,
$\bar\sigma^{\dot\alpha\beta\:\nu}\equiv (1,-\vec{\sigma})$ (here
$\vec{\sigma}$ are the usual Pauli matrices), we rewrite
Eq.~(\ref{eq:lagr4}) as
\begin{eqnarray}
  \label{eq:lagr}
  {\cal L}_{EM} = -\frac{1}{2}\mu_{a b} (\chi^\alpha)_a
  (\sigma^{\mu\nu})_{\alpha}^{\phantom{\alpha}\beta}
  (\chi_\beta)_b F_{\mu\nu} + {\rm h.c.}
\end{eqnarray}
This form makes the following important physical properties of the
interaction manifest:
\begin{itemize}
\item \emph{The interaction couples the neutrino and antineutrino
    states.} Notice in Eq.~(\ref{eq:lagr}) the spinors are either both
  ``undotted'' or both ``dotted''. Each ``undotted'' spinor destroys a
  neutrino state or creates an antineutrino state. 
\item  \emph{No flavor-diagonal terms.} Since spinors
anticommute, the $a=b$ terms in Eq.~(\ref{eq:lagr}) vanish
identically, and hence Majorana neutrinos cannot have flavor-diagonal
magnetic moments \cite{Valle1981,Okun1986}. This can be easily seen,
for example, using the definition of the
$(\sigma^{\mu\nu})_{\alpha}^{\phantom{\alpha}\beta}$ matrix given
above and the identity
\begin{equation}
  \label{eq:WB1}
  \chi \sigma^\mu \bar\sigma^\nu \psi = \psi \sigma^\nu \bar\sigma^\mu \chi
\end{equation}
The \emph{transition}
($a\neq b$) moments, however -- which lead to a simultaneous spin and
flavor change, {\it i.e.}, $\nu_e\rightarrow\bar\nu_{\mu,\tau}$ -- are
perfectly allowed.
\item \emph{The interaction with the $z$ (longitudinal) component of
    the magnetic field vanishes identically, while the interaction with the
    transverse components $\vec{B}_\perp=(\vec{B}_x,\vec{B}_y)$ is
    nonzero.}  One of the two spinors in Eq.~(\ref{eq:lagr}) has a
  raised index. Denoting it by $\tilde\chi\equiv
  \chi^{\alpha}=\epsilon^{\alpha\beta}\chi_\beta$, we write
  Eq.~(\ref{eq:lagr}) in the three-dimensional notation as
\begin{eqnarray}
  \label{eq:lagr3d}
  {\cal L}_{EM} = -\mu_{a b} (\tilde\chi)_a
  \vec{\sigma} (\nu)_b
  (\vec{B}-i\vec{E}) + {\rm h.c.}
\end{eqnarray}
Plugging in $\chi_a\propto\chi_b\propto(0\;1)^T$ -- which neglects
terms of order $m/E_\nu$ -- we find that the
interaction picks the $12$ component of the matrix $\sigma$, i.e., $1$
for $\sigma_x$, $-i$ for $\sigma_y$, and $0$ for $\sigma_z$.

The same conclusion can be reached by boosting the magnetic fields to
the neutrino rest frame.
\item The real part of $\mu_{a b}$ corresponds to the magnetic dipole
  moment and the imaginary part to the electric dipole moment. An
  ultrarelativistic neutrino precesses in an external magnetic field
  if \emph{either} the magnetic or electric dipole moments are
  non-zero \cite{Okun1986short,Okun1986}. Following the convention, we
  will henceforth use the term ``magnetic moment" to refer to both
  possibilities.
\end{itemize}

All these properties are well-known and are derived here for completeness.

\subsection{Size of the transition moment}\label{sect:directbounds}

The laboratory bounds on the neutrino magnetic (transition) moment
come from measuring the cross sections of $\nu e^-$ or $\bar\nu e^-$
scattering in nearly forward direction. The recent bound for the
interaction involving the electron antineutrino is $\mu_e < 1.0 \times
10^{-10}\mu_B$ at the 90\% confidence level \cite{munu2003}, where
$\mu_B \equiv e/(2 m_e)$ is the Bohr magneton ($m_e$ is the electron
mass, $e$ is its charge). Stronger bounds, $\mu \lesssim 3 \times
10^{-12}\mu_B$, exist from astrophysical considerations, particularly
from the study of red giant populations in globular clusters
\cite{Raffeltbound}.  Larger values of the transition moment would
provide an additional cooling mechanism and change the red giant core
mass at helium flash beyond what is observationally allowed.

Even allowing for new inputs (such as updated distances to globular
clusters) or refinements in the stellar models, it appears extremely
unlikely that the red giant bound could be weakened by more than a
factor of a few. Correspondingly, as a very conservative bound, we
will take $\mu < 10^{-11}\mu_B$ in the subsequent analysis.

From the theoretical point of view, the neutrino does possess a
nonzero transition moment, already in the SM. It is induced by loop
effects: as the neutrino briefly dissociates into virtual charged
particles ($W$ and a charged lepton $l$), those constituent particles
couple to the photon. Because the operator in Eq.~(\ref{eq:lagr})
requires a helicity flip, and because in the SM the $W$ boson couples
only to left-handed fields, the mass insertion has to be put on the
neutrino line. The resulting moment is proportional to the neutrino
mass and hence is highly suppressed, $\mu_\nu \sim e G_F m_\nu \sim
10^{-19} \mu_B (m_\nu/ eV)$. In possible extensions of the SM,
however, this need not be the case, and the effect can be proportional
to the mass of the charged lepton running in the loop.  For example,
in the left-right symmetric model one obtains $\mu_\nu \sim e G_F m_l
\sin 2\eta$ \cite{Kim1976}, where $\eta$ is the left-right mixing
parameter in the model.

\subsection{Solar magnetic fields}\label{sect:magnfields}

\subsubsection{Convective Zone}

The CZ of the Sun is known to host significant magnetic fields, which manifest
themselves in a variety of effects on the solar surface (sunspots,
flares, prominences, etc). These fields are created and destroyed
during each solar cycle by a combination of convective motions and
differential rotation. Detailed understanding of the solar cycle is
still an active subject of research. Nevertheless, the following
general features of the CZ fields are well established:
\begin{itemize}
 \item The magnetic field observed in sunspots
 is several kG, exceeding the
average surface field strength by three orders of magnitude.
Sunspots usually come in pairs of opposite polarity and are
thought to be manifestations of large-scale magnetic structures
residing inside the CZ.
 \item The total toroidal flux in the CZ at
 sunspot maximum can be estimated from the total flux that emerges
 on the surface during the solar cycle, which is around
 $2\times 10^{25}$ Mx \cite{GallowayWeiss}. Since the same flux tube may
 be emerging more
 than once, $10^{24}$ Mx has been argued to be the likely value
 \cite{GallowayWeiss}.
 Averaged over the CZ, this gives a field of several kG.
\item The turbulent equipartition value for the magnetic field at the
  base of the CZ can be expressed in terms of the density $\rho$, the
  solar luminosity $L_\odot$, and the distance to the center of the Sun
  $r$: $B_{\rm eq}\sim \rho^{1/6} L_\odot^{1/3} r^{-2/3} \sim 10$ kG
  (e.g., \cite{ourmagnfit}).
\item Modern models of the CZ fields argue that fields on the order of
  100 kG may exist in the CZ. For the present discussion, it is
  important to realize that such strong fields, if they exist, can
  occupy only a small fraction of the CZ volume, such as the
  \emph{thin} shear layer near the base of the CZ (`tachocline'), or
  be localized in isolated flux tubes
  \cite{fanliving}.  To have even half of the CZ filled with such
  strong fields would be contrary to the flux arguments mentioned
  above, as well as basic energetics considerations.
\end{itemize}

For an in-depth review of the CZ fields and further references, the
reader is urged to consult an excellent review \cite{fanliving}.  For
the status of the present-day simulation efforts see, e.g.,
\cite{toomre}.

\subsubsection{Radiative Zone} \label{sect:fields}

Unlike the CZ, the RZ has no internal motions to provide an active
mechanism to generate magnetic fields. On the other hand, the
conductivity of the RZ plasma is very high, which opens up the
possibility that the RZ could support large-scale magnetic fields left
over from the early stages of the solar evolution. Let us consider
both of these points in turn.

The evidence for the lack of any large-scale mixing motions in the RZ
comes from several sources. First, helioseismological studies show
that, unlike the CZ, the RZ rotates as a solid body. (This fact by
itself, incidentally, provides a hint that at least a weak poloidal
field is present in the RZ \cite{GoughMcIntyre}). Second, solar model
calculations show that the stratification (entropy gradient) in the RZ
interior is quite strong, precluding large radial motions. Finally, a
very impressive direct confirmation of the lack of radial mixing is
provided by the observed Beryllium abundance at the solar surface
\cite{Parker1984}. Be is destroyed at $T=3\times 10^6$ K, which is the
temperature at $r=0.61 R_\odot$, $R_\odot$ being the solar radius,
$6.96\times 10^5$ km. Since Beryllium observed at the solar surface is not
noticeably depleted relative to its primordial abundance, the material
at $r\sim 0.7 R_\odot$ (on the bottom of the CZ) and the material at
$r\sim 0.6 R_\odot$ have not been mixed during the lifetime of the Sun
(except for a brief period during the pre-main sequence convection
stage).

While a large-scale magnetic field in the RZ is not being presently
generated, it is also not destroyed by Ohmic decay over the solar
lifetime, 4.7 billion years. The Ohmic decay time for large-scale
field configurations is in the billions of years \cite{Cowling1945},
in fact for the lowest toroidal mode it is about 24 billion years
\cite{magnapj}. Over time, a primordial RZ field has been invoked by
many authors to explain a variety of effects: from the solar neutrino
problem \cite{Bartenwerfer,Chitre} to the so-called Princeton solar
oblateness measurements \cite{Dicke}.

What is important for the present analysis is that the field in the
RZ, if it exists, must be (i) smooth and (ii) bounded in strength
\cite{magnapj}. Smoothness follows from the fact that, because of
Ohmic decay, any small-scale magnetic field ($l\lesssim R_\odot/10 -
R_\odot/20$) features would have decayed away over the lifetime of the
Sun. This ensures the adiabatic character of the neutrino evolution
throughout the bulk of the CZ. The bounds on the field strength come
from several independent arguments, such as the measurements of the
solar oblateness and the stability analysis of field configurations.
It follows that the amplitude of the large scale toroidal field cannot
exceed $5-7$ MG \cite{magnapj} \footnote{See also \cite{Antia2002}
  where some of the same arguments are mentioned.}. Combined with the
bounds on the size of the neutrino transition moment, this limits the
value of the magnetic term (see below) in the neutrino oscillation
Hamiltonian.

\subsection{Oscillation Hamiltonian}\label{sect:oscHamiltonian}

Since flavor oscillations change the neutrino (and antineutrino)
flavor and the magnetic interactions convert neutrinos into
antineutrinos, the evolution of the neutrino state is, in general,
governed by a $6\times 6$ Hamiltonian matrix, with the neutrinos and
antineutrinos of all three flavors. For the purpose of studying the
spin flip in the RZ, the problem can be reduced to a $4\times 4$ case
\footnote{Since the atmospheric neutrino measurements indicate large
  mixing in the $\nu_\mu-\nu_\tau$ sector, solar neutrinos actually
  oscillate between $\nu_e$ and a certain superposition state of
  $\nu_\mu$ and $\nu_\tau$. To stay with the tradition, we denote the
  latter state by $\nu_\mu$ (and the corresponding superposition
  antineutrino state by $\bar\nu_\mu$).}. In the basis
$(\nu_e,\nu_\mu,\bar\nu_e,\bar\nu_\mu)$, the Hamiltonian is
\cite{LimMarciano}
\begin{eqnarray}
\label{eq:H}
 H= \left(\begin{array}{cc}
 H_\nu & (B_x-i B_y)M^\dagger \\
 (B_x+i B_y)M &  H_{\bar\nu}
\end{array}\right),
\end{eqnarray}
where $B_{x,y}$ are the transverse components of the magnetic field
and the $2\times 2$ submatrices are given by
\begin{eqnarray}
\label{eq:H1}
M&=&\left(\begin{array}{cc}
  0& -\mu_{e\mu} \\
 \mu_{e\mu} &  0
\end{array}\right),\\
\label{eq:H2}
 H_{\nu}&=&\left(\begin{array}{cc}
  -\Delta\cos 2\theta + A_e& \Delta\sin 2\theta \\
 \Delta\sin 2\theta &  \Delta\cos 2\theta + A_\mu
\end{array}\right),\\
\label{eq:H3}
 H_{\bar\nu}&=&\left(\begin{array}{cc}
  -\Delta\cos 2\theta - A_e& \Delta\sin 2\theta \\
 \Delta\sin 2\theta &  \Delta\cos 2\theta - A_\mu
\end{array}\right).
\end{eqnarray}
Here $\Delta\equiv\Delta m^2/4 E_\nu$,
$A_e\equiv\sqrt{2}G_F(n_e-n_n/2)$ and $A_\mu\equiv\sqrt{2}G_F
(-n_n/2)$.  $\Delta m^2$ is the neutrino mass-squared splitting,
$E_\nu$ is its energy, and $n_e$ and $n_n$ are the electron and
neutron number densities.

The physical reason for the $6\rightarrow 4$ reduction is that the
large atmospheric neutrino splitting effectively decouples two of the
six states in the case of the smooth magnetic field. It should be
noted, however, that the ``noisy'' field in the CZ enables transitions
between states separated by the large splitting. Consequently, in the
CZ one may need to consider the full six-neutrino problem
\cite{Guzzo2005}.

\section{Radiative zone}\label{sect:rz}

\subsection{Conventional approach}\label{sect:rz_conventional}

Historically, in discussions of spin-flavor precession (SFP) in the
Sun, special attention has been given to the ``SFP resonance" region
\cite{Akhmedov1988,LimMarciano}, defined as the region where two of
the \emph{diagonal} terms in the Hamiltonian (\ref{eq:H}-\ref{eq:H3}),
namely those corresponding to the $ee$ and $\bar\mu\bar\mu$ entries,
become equal. The corresponding resonance condition is
\begin{equation}\label{eq:res_diag}
    \sqrt{2}G_F(n_e-n_n) = \frac{\Delta m^2}{2 E_\nu} \cos 2\theta.
\end{equation}
Compared to the traditional ``flavor resonance" condition, defined
as the equality of the $ee$ and $\mu\mu$ diagonal entries in
(\ref{eq:H}-\ref{eq:H3})),
\begin{equation}\label{eq:res_diag_MSW}
    \sqrt{2}G_F n_e = \frac{\Delta m^2}{2 E_\nu} \cos 2\theta,
\end{equation}
condition (\ref{eq:res_diag}) is satisfied at a somewhat higher
density in the Sun.

When the resonant SFP mechanism was introduced back in the late
1980's, the values of the oscillation parameters were essentially
unknown. In contrast, nowadays the solar and KamLAND experiments
have told us that $\Delta m^2 \sim 8 \times 10^{-5}$ eV$^2$ and
$\tan^2\theta\sim 0.4$ \cite{KamLANDspectrum}. Two implications of
this have been recognized: (i) if the process of SFP operates in
the Sun, it must be combined with the ``ordinary" flavor
oscillations
\cite{AkhmedovPulido2002,Chauhanetalcore2003,rashbaprl,BalantekinVolpe},
and (ii) the SFP resonance condition, as given in
Eq.~(\ref{eq:res_diag}), is satisfied for $^8$B neutrinos in the
RZ.

\begin{figure}[htbp]
  \centering
  \includegraphics[width=0.47\textwidth]{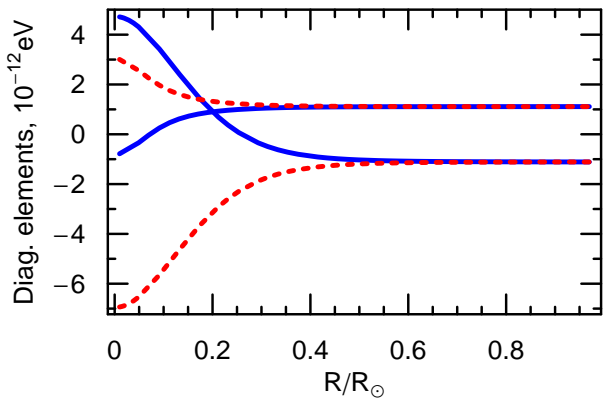}
  \caption{The diagonal elements of the Hamiltonian
    (\ref{eq:H}-\ref{eq:H3}) for $\Delta m^2=8\times 10^{-5}$ eV$^2$,
    $E_\nu=9$ MeV. The solid/dashed lines correspond to the
    neutrino/antineutrino entries.}
  \label{fig:levels_diag}
\end{figure}

The second point is illustrated in Fig.~\ref{fig:levels_diag}, where the
four diagonal elements of the Hamiltonian are plotted as a function of
the distance from the solar center. Here and later in this paper, the
BP04 solar model \cite{bahcallwebpage} is used.

In most of the recent studies of the RZ spin-flip
\cite{AkhmedovPulido2002,Chauhanetalcore2003,BalantekinVolpe}, the two
processes, flavor oscillations and spin-flavor precession, are treated
as occurring in spatially separated regions. The resulting probability
of $\nu_e\rightarrow\bar\nu_e$ conversion is then given as a product
\begin{equation}\label{eq:product}
    P(\nu_e\rightarrow\bar\nu_e) = P(\nu_e\rightarrow\bar\nu_\mu)
    \times P(\bar\nu_\mu\rightarrow\bar\nu_e).
\end{equation}

A notable exception is Ref.~\cite{rashbaprl}, where it is noted
that the evolution in the case of the RZ does not decouple into two
separate problems. The numerical conclusions  presented in Fig.~2 of
that paper, however, are quite different from ours, as discussed later.

\subsection{Large mixing and the disappearance of the resonance}
\label{sect:rz_nores}

The above treatment relies on the notion of resonance, as given by
Eq.~(\ref{eq:res_diag}). Fundamentally, this notion
comes from the analysis of a two-level system, and even there it
applies only in the limit of small mixing \cite{myres,myresproc}.
We wish to examine whether it can be applied to the four-level
system at hand.

Let us first briefly review the physics behind the standard
small-angle MSW effect \cite{W77,MS}. Suppose we are given a two-level
neutrino system with a Hamiltonian
\begin{equation}\label{eq:H2level}
    H_{2\nu} = \left(\begin{array}{cc}
  -D & d \\
 d &  D
\end{array}\right).
\end{equation}
If $D \gg d$, flavor oscillations are suppressed, as observed already
in \cite{W77}. Indeed, the mixing angle is given by $\tan 2\theta =
d/D$ and the oscillation amplitude is proportional to $\sin^2
2\theta$. The physical idea behind the resonance is that this
suppression is lifted if the diagonal splitting becomes small in some
region along the neutrino trajectory, as a result of the crossing of
the levels. Suppose $|D(l)|\lesssim |d(l)|$ over a distance $\delta
l$.  Then, inside this interval, $d$ becomes \emph{the dominant part
  of the Hamiltonian} and hence drives oscillations. If $\delta l$ is
sufficiently large, it may be possible to obtain large flavor
conversion \footnote{The conversion probability in the resonance
  region depends on whether the cancellation persists ``long enough",
  {\it i.e.}, whether the oscillation length due to $d$ fits inside
  the interval $\delta l$.  This condition gives nothing but the usual
  adiabaticity parameter of the resonance \cite{myresproc}.}.

With this in mind, let us return to the Hamiltonian in
Eqs.~(\ref{eq:H}-\ref{eq:H3}). Now, the electron neutrino state, in
addition to being coupled to the $\bar\nu_\mu$ state via the magnetic
moment interaction, is also coupled to $\nu_\mu$ state via the flavor
oscillation term. A simple, but crucial observation is
the following: for large values of the mixing angle, even if the
diagonal $ee$ and $\bar\mu\bar\mu$ terms are equal, the Hamiltonian is
\emph{still not dominated} by the off-diagonal magnetic terms $\mu B$,
but by the off-diagonal flavor oscillation terms $\Delta \sin
2\theta$. Thus, there is no reason to expect that the condition in
Eq.~(\ref{eq:res_diag}) should automatically lead to large
$\nu_e\rightarrow\bar\nu_\mu$ conversion.

This complication is, however, easily removed by going to a basis
in which the large off-diagonal flavor oscillation terms are
absent. Consequently, we diagonalize $H_\nu$ and $H_{\bar\nu}$
given in Eqs.~(\ref{eq:H2}) and (\ref{eq:H3}) by performing
separate flavor rotations on neutrinos and antineutrinos. We find
\begin{widetext}
\begin{eqnarray}
\label{eq:Hmass}
H = \begin{pmatrix}
\Delta_{m1}  & 0 & \mu_{e\mu} B \sin(\theta_m-\bar\theta_m) & \mu_{e\mu} B \cos(\theta_m-\bar\theta_m) \\
0 & \Delta_{m2} & -\mu_{e\mu} B \cos(\theta_m-\bar\theta_m) &\mu_{e\mu} B \cos(\theta_m-\bar\theta_m) \\
\mu_{e\mu} B \sin(\theta_m-\bar\theta_m) & -\mu_{e\mu} B \cos(\theta_m-\bar\theta_m) & \bar\Delta_{m1} & 0\\
\mu_{e\mu} B \cos(\theta_m-\bar\theta_m) & \mu_{e\mu} B \sin(\theta_m-\bar\theta_m) & 0 & \bar\Delta_{m2}
\end{pmatrix}.
\end{eqnarray}
\end{widetext}

In Eq.~(\ref{eq:Hmass}), $\Delta_{m1}$, $\Delta_{m2}$,
$\bar\Delta_{m1}$ and $\bar\Delta_{m2}$ are the eigenvalues of the
oscillation Hamiltonian (\ref{eq:H}-\ref{eq:H3}) for vanishing
magnetic field,
\begin{widetext}
\begin{eqnarray}
    \label{eq:deltami}
    \Delta_{mi}&=& \frac{A_e + A_n}{2}
    \pm \left[\left(\Delta\cos 2\theta-\frac{A_e-A_n}{2}\right)^2+\Delta^2\sin^2
    2\theta\right]^{\frac{1}{2}},\\
    \label{eq:deltamibar}
\bar\Delta_{mi} &=& -\frac{A_e + A_n}{2}
    \pm \left[\left(\Delta\cos 2\theta+\frac{A_e-A_n}{2}\right)^2+\Delta^2\sin^2
    2\theta\right]^{\frac{1}{2}},
\end{eqnarray}
\end{widetext}
while $\theta_m$ and $\bar\theta_m$ are the rotation angles,
\begin{eqnarray}
    \tan 2\theta &=& \Delta\sin 2\theta/[\Delta\cos 2\theta-(A_e-A_n)/2],\\
    \tan 2\bar\theta  &=& \Delta\sin 2\theta/[\Delta\cos 2\theta+(A_e-A_n)/2].
    \label{eq:mixingangle}
\end{eqnarray}
The barred quantities refer to antineutrinos.

Let us first discuss the general properties of these equations.
When the matter potentials are negligible, $\Delta_{m1}$ and
$\bar\Delta_{m1}$ both approach $-\Delta$, while $\Delta_{m2}$ and
$\bar\Delta_{m2}$ both approach $\Delta$. Likewise, in the same limit
the rotation angles $\theta_m$ and $\bar\theta_m$ both approach the
vacuum value $\theta$. Notice that in vacuum the structure of the
off-diagonal magnetic interactions in the mass basis is the same as in
the flavor basis.  This structure is that of an antisymmetric tensor,
Eq.~(\ref{eq:H1}), which is invariant under an $SU(2)$ rotation
applied to both indices. More generally, this is a manifestation of
the fact that the operator in (\ref{eq:lagr}) cannot convert a
Majorana neutrino state in vacuum into its own antiparticle, whether
this state is a mass eigenstate or a flavor eigenstate.  In
sufficiently dense matter, however, $\theta_m$ and $\bar\theta_m$ are,
in general, different, because the matter terms for neutrinos and
antineutrinos have opposite signs.  Thus, in matter, direct
transitions between all four states become allowed \footnote{An
  extremely illuminating analysis of this in perturbation theory can
  be found in the classical work on the subject \cite{Akhmedov:1989}.
  There, it was observed that the perturbative amplitudes for
  $\nu_e\rightarrow\nu_\mu\rightarrow\bar\nu_e$ and
  $\nu_e\rightarrow\bar\nu_\mu\rightarrow\bar\nu_e$ have opposite
  signs and hence tend to cancel. The cancellation is incomplete if
  one of the channels is enhanced with respect to the other one by
  matter effects. }.

For the energies at which KamLAND is sensitive to solar antineutrinos,
$E_\nu\gtrsim 8.3$ MeV, the magnitude of $2\Delta$ is of order $\sim 8
\times 10^{-5} /(2 \times 10^7) \sim 4 \times 10^{-12}$ eV. By
comparison, the value of the off-diagonal magnetic terms is at most of
order $10^{-11} \mu_B \times 5$ MG $\sim 3 \times 10^{-13}$ eV, or at
least an order of magnitude smaller. This makes the analysis in the
mass basis quite straightforward. Clearly, transitions between any two
states with diagonal splitting of order $10^{-12}-10^{-11}$ eV will be
suppressed. Conversions between a pair of states with smaller diagonal
splitting may take place, provided that $\theta_m$ and $\bar\theta_m$
are sufficiently different, {\it i.e.}, the matter potential is
sufficiently large.

\begin{figure}[htbp]
  \centering
  \includegraphics[width=0.47\textwidth]{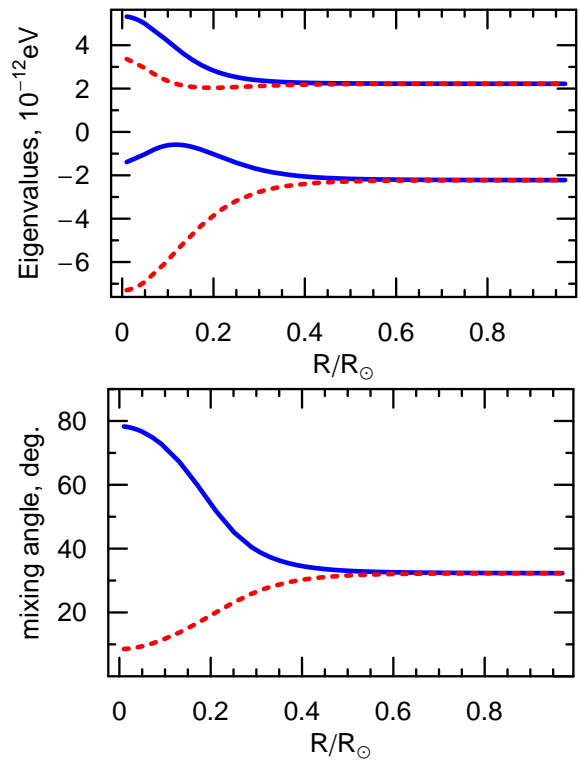}
  \caption{ The diagonal elements of the Hamiltonian
    (\ref{eq:Hmass}) (top panel) and the mixing angles $\theta_m$ and
    $\bar\theta_m$ (bottom panel) for $\Delta m^2=8\times 10^{-5}$
    eV$^2$, $\tan^2\theta=0.4$, $E_\nu=9$ MeV. The solid/dashed lines
    correspond to the neutrino/antineutrino levels.}
  \label{fig:eigen_mixing}
\end{figure}

In the top panel of Fig.~\ref{fig:eigen_mixing} we show the diagonal
elements of the Hamiltonian Eq.~(\ref{eq:Hmass}) for the best-fit
values of the oscillation parameters. Obviously, their behavior is
quite different from the diagonal elements in the flavor basis,
plotted in Fig.~\ref{fig:levels_diag}. In particular, in
Fig.~\ref{fig:eigen_mixing} there is no level crossing at $R\simeq
0.18R_\odot$. \emph{The mass eigenvalues do not intersect anywhere
  inside the Sun.}

Notice that the top two levels in Fig.~\ref{fig:eigen_mixing} get
close to each other for $r\gtrsim 0.3-0.4 R_\odot$. The difference
$\theta_m-\bar\theta_m$ there is small, but non-zero, and \textit{a
  priori} could drive a transition between these two levels. The
possibility of this transition requires a more detailed study. 

The answer comes from considering the \emph{adiabaticity} of the
neutrino evolution. Consider an idealized problem: a neutrino mass
eigenstate moving through a magnetic field which is adiabatically
turned on and then turned off.  The neutrino will be partially
converted into the corresponding antineutrino state \emph{while inside
  the field region}, but upon exiting the region will revert back to
being fully a neutrino.

This idealized description applies in our case: (i) At the production
point in the core, for the best-fit KamLAND mass splitting $\Delta
m^2\sim 8\times 10^{-5}$ eV$^2$, the diagonal splitting dominates,
suppressing the effects of the magnetic field. The neutrino is
produced predominately in the heavy eigenstate (the top line in the
upper panel of Fig.~\ref{fig:eigen_mixing}). (ii) In the region $r\sim
0.2-0.5 R_\odot$, the magnetic term drives partial
$\nu\leftrightarrow\bar\nu$ conversion (transitions between the top
two states in the top panel of Fig.~\ref{fig:eigen_mixing}). (iii) As
the effects of the magnetic term are decreased (mainly due to the
alignment of $\theta_m$ and $\bar\theta_m$ at smaller densities), the
$\nu\leftrightarrow\bar\nu$ conversion is largely undone. The
adiabaticity through the bulk of the RZ zone is ensured by the physics
of the RZ fields: as mentioned in Sec.~\ref{sect:fields}, any
structures smaller than $ R_\odot/10 - R_\odot/20$ must have decayed
away over the lifetime of the Sun.

\begin{figure}[htbp]
  \centering
  \includegraphics[width=0.47\textwidth]{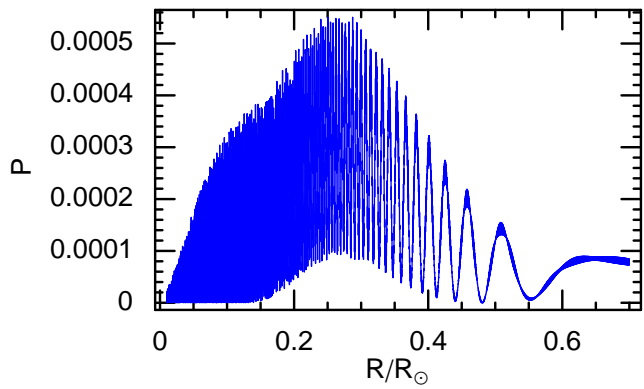}
  \caption{A typical case of the neutrino evolution, for the value of the
    RZ field near the limit. Shown is the probability of
    finding a neutrino in the $\bar\nu_e$ state as a function of the
    distance from the solar center (in units of the solar radius), for
  $\Delta m^2=8\times 10^{-5}$ eV$^2$, $E_\nu=10$ MeV, $\theta=\pi/6$,
$B_{\rm max}=5$ MG.}
  \label{fig:evolution}
\end{figure}

This behavior is illustrated in Fig.~\ref{fig:evolution}, which shows
that the electron antineutrino fraction $P$ indeed first grows, but
then decreases to unobservable levels, as the effect of the magnetic
term decreases in the outer part of the RZ. The residual nonzero value
of $P$ is due to the violation of adiabaticity near the top of the RZ.

In summary, except in the limit of small mixing angle,
Eq.~(\ref{eq:res_diag}) is not the right condition for describing
resonant neutrino-antineutrino conversion.  A correct physical
criterion should be based on comparing the diagonal elements \emph{in
  the mass basis}, not in the flavor basis. This ``mass basis
level-crossing'' would be indeed realized in the Sun, had the value of
the mixing angle been small\footnote{This was how the solution
  described in \cite{ourmagnfit} was constructed.  Notice that there
  it was essential that the mixing angle was small.}. For the large
measured value of the mixing angle, however, the level-crossing is
simply absent and the efficiency of the neutrino-antineutrino
conversion is very low.

\subsection{Large mixing and level crossing: analytical treatment}
\label{sect:rz_analyt}

Let us now derive the general analytical condition for having resonant
neutrino-antineutrino conversion.
Referring to Fig.~\ref{fig:eigen_mixing}, we see that, {\it a priori},
either the two light levels or the two heavy levels could
intersect. The corresponding analytical conditions for these two
possibilities are 
\begin{equation}
    \label{eq:cond}
   \Delta_{m2}=\bar\Delta_{m2},
\end{equation}
and
\begin{equation}
    \label{eq:condm1}
   \Delta_{m1}=\bar\Delta_{m1},
\end{equation}
where the indices ``1'' and ``2" refer to the light and heavy states
respectively (the ``-" and ``+" signs in
Eqs.~(\ref{eq:deltami},\ref{eq:deltamibar})).

Written out explicitly, the above two conditions read
\begin{widetext}
\begin{eqnarray}
  \label{eq:cond_a}
  a\sqrt{2}G_F(n_e - n_n) =  
\left[\left(\sqrt{2}G_F n_e/2+\Delta\cos 2\theta\right)^2+\Delta^2\sin^2
    2\theta\right]^{\frac{1}{2}}
-\left[\left(\sqrt{2}G_F n_e/2-\Delta\cos 2\theta\right)^2+\Delta^2\sin^2
    2\theta\right]^{\frac{1}{2}},
\end{eqnarray}
\end{widetext}
where $a$ is $+1$ for the condition in Eq.~(\ref{eq:cond}) and $-1$
for the condition in Eq.~(\ref{eq:condm1}). Since in the Sun $n_e -
n_n > 0$ always, the above condition can be satisfied only for $a=-1$ for
$\theta>\pi/4$ and only for $a=+1$ for $\theta<\pi/4$. In other words,
in the ``dark side'' \cite{darkside,vacuummsw} only the two light
levels can intersect, while in the ``light side'', as preferred by the
data, only the two heavy levels can.


\begin{figure*}[htbp]
  \centering
  \includegraphics[width=0.87\textwidth]{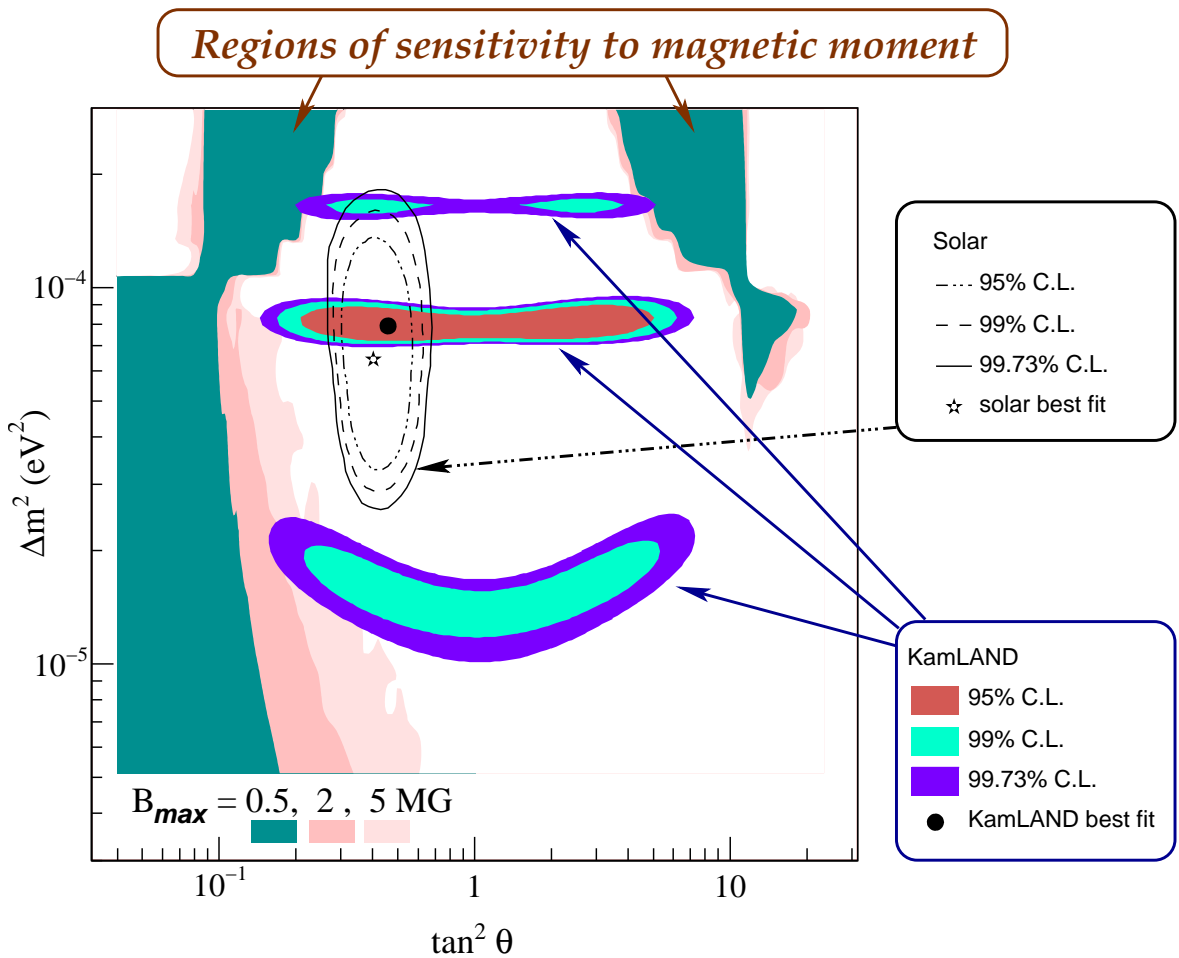}
  \caption{The regions of the oscillation parameter space where
    one may expect the electron antineutrino flux above the KamLAND
    bound \cite{KamLANDnubar} (three different shadings
    correspond to three different normalizations of the magnetic
    field). A very optimistic value of the transition moment, $\mu=
    1\times 10^{-11}\mu_B$, was taken. Also shown are the regions
    allowed by the analysis of the KamLAND data \cite{KamLANDspectrum}
    (shaded regions in the middle) and the region allowed by the solar
    data (unfilled contours).}
  \label{fig:scan}
\end{figure*}

With this in mind, let us solve Eq.~(\ref{eq:cond_a}). We find,
in addition to the trivial vacuum solution, $n_e=n_n=0$, a
solution
\begin{equation}\label{eq:cond_solution}
    \frac{\Delta m^2}{2 E_\nu} = \sqrt{2}G_F(n_e-n_n)
    \sqrt{\frac{n_e^2-(n_e-n_n)^2}{n_e^2 \cos^2
    2\theta-(n_e-n_n)^2}}.
\end{equation}
This is the desired general condition for having resonant antineutrino
production in the solar magnetic field. Once again, for
$\theta<\pi/4$, it solves Eq.~(\ref{eq:cond}), while in the opposite
case it solves Eq.~(\ref{eq:condm1}). (It is also worth reminding here
that (\ref{eq:cond_solution}) has been derived assuming the neutrino
is a Majorana particle.)

We indeed see that Eq.~(\ref{eq:cond_solution}) and the conventionally used
Eq.~(\ref{eq:res_diag}) agree only for $\theta=0$ and are
substantially different for large $\theta$. In particular, as $\theta$
is increased beyond
\begin{equation}\label{eq:analyt_theta}
    |\cos 2\theta_{\rm crit}| \simeq (1-n_n/n_e),
\end{equation}
the level crossing as given by Eq.~(\ref{eq:cond_solution}) 
disappears. It can be said that there is a ``topological'' difference
between the cases of small and large mixing and
Eq.~(\ref{eq:analyt_theta}) gives the angle at which the topology changes.

The ratio $n_n/n_e$ varies with the distance to the center of the
Sun. Correspondingly, the numerical value of $\theta_{\rm crit}$
for which the disappearance takes place varies with $\Delta$
(which controls the location in the Sun where the level crossing
could have happened). In the central region, $n_n/n_e \simeq 1/2$,
while for the outer regions $n_n/n_e \simeq 1/6$. The
corresponding variation in $\theta_{\rm crit}$ is
\begin{equation}\label{eq:thetacrit}
    \tan^2 \theta_{\rm crit} \sim (0.09 - 0.33).
\end{equation}

Let us finally discuss when each of the
level-crossings, (\ref{eq:cond}) and (\ref{eq:condm1}), can affect the
neutrino evolution. The answer depends on the probability of finding
the initial electron neutrino in each of the two neutrino mass
eigenstates. This probability, in turn, depends on the flavor mixing
angle at the production point. If the matter potential $\sqrt{2} G_F
n_e$ dominates over the vacuum term $\Delta m^2/(2 E_\nu)$, the mixing
angle at production approaches $\pi/2$ and the neutrino is
produced almost entirely in the heavy neutrino mass eigenstate. In
the opposite limit, the neutrino is produced in the superposition of
both mass eigenstates, with the probabilities $\cos^2 \theta$ and
$\sin^2 \theta$.

The rough boundary between these two regimes for the $\sim$10 MeV $^8$B
neutrinos lies near $\Delta m^2 \sim 10^{-4}$ eV$^2$: for smaller
$\Delta m^2$ including the best-fit region of $\Delta m^2 \sim
8\times 10^{-5}$ eV$^2$ only the condition in
Eq.~(\ref{eq:cond}) matters, while in the opposite case both
Eq.~(\ref{eq:cond}) and Eq.~(\ref{eq:condm1}) do. For $\Delta m^2
\gtrsim 10^{-4}$ eV$^2$ we can thus expect sizable antineutrino production
both in the ``light'' and ``dark'' sides, while for $\Delta m^2
\ll 10^{-4}$ eV$^2$ the effect can happen only in the light side.

\subsection{Large mixing and $\bar\nu_e$ production: numerical treatment}
\label{sect:rz_scan}

To confirm this analysis, we have carried out a numerical scan of
the oscillation parameter space, in the range $0.5 \times
10^{-5}<\Delta m^2 <3\times 10^{-5}$ eV$^2$. The calculation was
carried out for the realistic numerical solar density and chemical
composition profiles \cite{bahcallwebpage}. For the magnetic field
profile, we have taken the lowest toroidal Ohmic eigenmode, as
described in \cite{ourmagnfit} and in detail in \cite{magnapj}.
While higher eigenmodes could also be present, the choice of the
lowest mode is expected to capture the essential physics: the
field is axisymmetric, smooth, and vanishes at the RZ/CZ boundary.
The only free parameter describing the field in this case is the
overall mode normalization. We have performed scans for three
different (highly optimistic) normalization values of the magnetic
field, $B_{\rm max}= 0.5, 2, 5$ MG, also taking a highly
optimistic value of the transition moment, $\mu= 1\times
10^{-11}\mu_B$.

The results of the scan are shown in Fig.~\ref{fig:scan}. The three
colored regions on the left side (small mixing) indicate the regions
where the $\bar\nu_e$ flux for the chosen normalization values of the
magnetic field would exceed the KamLAND bound. The corresponding
regions in the ``dark" side, $\theta>\pi/4$,
are also shown. All the regions agree well with the analytical
analysis of the previous Subsection. 

The Figure also shows the allowed regions from the KamLAND
analysis of their reactor antineutrino spectrum
\cite{KamLANDspectrum}, as well as the region that gives a fit to the
solar neutrino data (unfilled contours). The best-fit (``LMA-I'')
region lies at the intersection of the solar and KamLAND regions and,
as can be clearly seen, has no overlap with the regions where the
$\bar\nu_e$ flux would be measurable.

We note that a similar conclusion was reached in
\cite{AkhmedovPulido2002}, using, however, an analytical criterion
that differs from Eq.~(\ref{eq:analyt_theta}). The relatively
large $\bar\nu_e$ flux cited in that work was obtained by assuming
the field strength of 50 MG at the neutrino production point,
which was only later shown to be not allowed \cite{magnapj}.

\section{Convective zone}\label{sect:cz}

Compared to the case of the RZ, the case of the CZ requires a
different treatment. The difference has to do with the turbulent
nature of the CZ magnetic fields. Below, for completeness, we briefly
review the relevant physics.

The main idea is that random variations \cite{Loreti1994,Burgess1996}
of the magnetic field on scales equal to, or smaller than, the
neutrino oscillation length lessen the suppression of the oscillations
by a large diagonal mass splitting. While the neutrino follows slow
variations with scales longer than the oscillation length
$\lambda_{\rm osc} \sim 1/\Delta$ adiabatically, for shorter
variations the adiabaticity condition is violated. The net result of
the small-scale ``noise" is that the neutrino state vector performs a
``random walk" in the oscillation space, gradually receding from its
original position.

To be specific, consider a simple model in which the magnetic field
changes on scales of $\lambda_{\rm corr} \lesssim \lambda_{\rm osc}$.
The probability of spin flip upon traversing the first correlation
cell, $\lambda_{\rm corr}$, is $(\mu B \lambda_{\rm corr})^2$ (simple
vacuum oscillation on scales smaller than the oscillation length).
Since the fields in different $\lambda_{\rm corr}$ intervals are
uncorrelated, as the neutrino traverses many domains the
probabilities, not amplitudes, add up. For solar neutrinos, the
experimental bound on $\bar\nu_e$ flux implies $P\ll 1$. So long as
this is satisfied, the probability of spin flip after traveling the
distance $L$ can be written as $P\sim(\mu B \lambda_{\rm corr})^2
L/\lambda_{\rm corr}$.

For slow variations with $\lambda_{\rm corr} \gg \lambda_{\rm osc}$,
as already mentioned, the evolution is adiabatic and the conversion
probability is exponentially suppressed. One can also imagine a case
when the field domains are longer than $\lambda_{\rm osc}$, but have
sharp boundaries. In such a case, the conversion probability is $P\sim
\sin^2 2\theta L/\lambda_{\rm corr} \sim (\mu B/\Delta)^2
L/\lambda_{\rm corr}$ We see that in any case the conversion is most
efficient for $\lambda_{\rm corr} \sim \lambda_{\rm osc} \sim
1/\Delta$, namely, $P_{\rm opt}\sim(\mu B)^2 L \lambda_{\rm osc}$.

Let us apply this to solar neutrinos. Once again, we are
interested in the high-energy $^8$B neutrinos, as these are the
neutrinos the KamLAND bound applies to. These neutrinos are
produced almost exclusively in the heavy mass eigenstate $\nu_2$
and then propagate adiabatically through the RZ. In the CZ, these
neutrinos encounter turbulent magnetic fields and, if the
transition moment is large, a small fraction of them gets
converted into the light antineutrino mass eigenstate $\bar\nu_1$.
The net result is the flux of neutrinos in the heavy eigenstate
with possibly a small admixture of the antineutrinos in the light
mass eigenstate. The light mass eigenstate, in turn, is detected
as an electron antineutrino with probability $\cos^2\theta$. The
relevant field fluctuations are those on scales of $\lambda_{\rm
  osc} \sim 2 \times 10$ MeV$/(8\times 10^{-5})$ eV$^2 \sim 3 \times
10^2$ km. We will denote them by $B_{\lambda_{\rm osc}}$. Overall, we
get
\begin{equation}\label{eq:CZP}
    P(\nu_e\rightarrow\bar\nu_e)\sim\cos^2\theta(\mu B_{\lambda_{\rm osc}})^2
    L\lambda_{\rm osc}.
\end{equation}

The oscillation parameters are known, $\cos^2\theta \sim 0.7$, $\Delta
\sim 10^{-11} \mbox{ eV} (10 \mbox{ eV}/ E_\nu)$. The extent of the
region of the strong random field can be estimated to be some fraction
of the CZ, say $L \sim 0.1 R_\odot$. Finally, the main issue is to
estimate the size of the amplitude of the small-scale noise
$B_{\lambda_{\rm osc}}$. Unfortunately, the theory and simulation
efforts are not yet at the level to give a robust prediction of
$B_{\lambda_{\rm osc}}$ (see \cite{toomre} for the state of the art on
the simulation front). We will next consider two models for the
magnetic field.

\subsection{Model of ``uniform'' turbulence}
\label{sect:cz_Kolmogorov}

A physically sensible estimate can be obtained
by assuming that (i) the field on the large scales is of the order of
its turbulent equipartition value; and (ii) that the small-scale noise
follows some kind of a scaling law
\begin{equation}\label{eq:Bscaling}
    B_{\lambda} \propto \lambda^\alpha,
\end{equation}
typical of turbulent systems. Such approach was adopted in
\cite{rashbaprl,rashbaprd} and, in our opinion, is a clear improvement
over naive models of delta-correlated noise.

The largest scale of the turbulence, at which energy pumping takes
place, is estimated to be of order the typical pressure scale height
in the CZ. In the inner part of the CZ, the pressure scale height is
of order $L_{\rm max}\sim 5\times 10^{4}$ km \cite{fanliving}.  This
number can be estimated directly from the solar model
\cite{bahcallwebpage}, and is also consistent with the value of the
mixing length $8\times 10^4$ km at the base of the CZ given in
\cite{zeldovich}.  Although the corresponding large-scale eddies
reside in the interior of the CZ, they manifest themselves on the
solar surface as \emph{supergranules} ($l\simeq 15\times 10^4$ km).
These should not be confused with surface granules ($l\simeq 10^3$ km)
whose size is dictated by the scale height near the surface.

The turbulent equipartition value for the large scale turbulent fields
near the bottom of the CZ is estimated as $B_{L_{\rm max}}\sim
\rho^{1/6} L_\odot^{1/3} r^{-2/3} \sim 10$ kG \cite{ourmagnfit}.

Putting the numbers together, we find
\begin{widetext}
\begin{equation}\label{eq:CZPnum}
    P(\nu_e\rightarrow\bar\nu_e)\sim 10^{-5}
    \left(\frac{\mu}{10^{-11} \mu_B} \right)^2
    \left(\frac{B_{L_{\rm max}}}{10 \mbox{ kG}}\right)^2
    \left(\frac{\lambda_{\rm osc}}{L_{\rm max}}\right)^{2\alpha-\frac{2}{3}}
    \left(\frac{L}{10^5 \mbox{ km}}\right)
    \left(\frac{E_\nu}{10 \mbox{ MeV}}\right)
    \left(\frac{8\times 10^{-5} \mbox{ eV}^2}{\Delta m^2}\right)
    \left(\frac{\cos^2\theta}{0.7}\right).
\end{equation}
\end{widetext}
This means that, for this field model, even with the most
optimistic values of the neutrino magnetic moment KamLAND will not
have a chance to see antineutrinos from the CZ spin flip unless its
sensitivity is improved by as least one order of magnitude.

\subsection{Model of isolated flux tubes}
\label{sect:cz_tubes}

The estimate just given uses a simple model for the turbulent magnetic
field. It is possible that the field in the CZ instead has a
``fibril'' nature \cite{fanliving}, i.e., is expelled by the
turbulence and combines in isolated flux tubes. It has been argued
that the total energy of the CZ (thermal + gravitational + magnetic)
is reduced by the fibril state of the magnetic field by avoiding the
magnetic inhibition of convection \cite{Parker1984b}. 

Let us then consider a model in which the neutrinos travel through
spatially separated flux bundles. The conversion probability in this
case depends on the transverse size of a bundle. Given that a
sunspot typically contains $10^{20}$ Mx \cite{GallowayWeiss,fanliving}
and assuming superequipartition field strength inside the bundle of
$\sim 100$ kG, one finds the transverse size of the bundle that is
actually close to optimal, $\sim 300$ km.  The thickness of the flux
rope comes out to be close to the oscillation length for the $^8$B
neutrinos, a remarkable coincidence indeed!  Since the total toroidal
flux in the CZ is estimated to be of the order of $10^{24}$ Mx
\cite{GallowayWeiss}, a given neutrino encounters only several of such
flux bundles. The resulting conversion probability in the most
optimistic case is then close to the KamLAND sensitivity bound, as can
be easily checked. 

\section{Conclusions}
\label{sect:conclusions}

We have shown that the large measured value of the flavor mixing angle
\emph{qualitatively} changes the treatment of the neutrino interaction
with the magnetic field in the radiative zone of the Sun. The physics
of the effect is most transparent in the basis of the mass
eigenstates, rather than in the flavor basis.
We have shown that in the mass basis, for the allowed values of the
oscillation parameters, there is no neutrino-antineutrino level
crossing and, as a result, no measurable antineutrino production in
the RZ.
A correct level crossing condition, Eq.~(\ref{eq:cond_solution}), was
derived, and a detailed numerical scan carried out to confirm
this conclusion. 

We emphasize that the recent determination of the neutrino oscillation
parameters is crucial for reaching this conclusion: for small flavor
mixing angle, it was possible to have large
$\nu_e\rightarrow\bar\nu_\mu$ conversion even with magnetic fields an
order of magnitude below the bound, $B\sim 0.5$ MG \cite{ourmagnfit}.
Moreover, as seen in Fig.~\ref{fig:scan}, even a relatively minor
shift of the best-fit point, from $\Delta m^2 \simeq 0.8 \times
10^{-4}$ eV$^2$ and $\theta\simeq 32^\circ$ to, {\it
  e.g.}, $\Delta m^2 \simeq 2 \times 10^{-4}$ eV$^2$ and $\theta\simeq
24^\circ$, would have made the resonant antineutrino production
possible. The field of solar neutrinos has truly entered the precision
measurement stage!

One may wonder if our analysis is completely self-consistent. After
all, the oscillation parameters we use are found assuming only the
flavor oscillations and neglecting possible magnetic
$\nu\leftrightarrow\bar\nu$ transitions. This, however, is entirely
justified, because the KamLAND bound on the $\bar\nu_e$ flux ensures
that the magnetic transitions can be present only at very small levels
and thus cannot affect the fit to the oscillation parameters.

In the convective zone, the solar neutrino propagate through turbulent
magnetic fields. Our estimate of the probability of the spin-flavor
flip in this case gives a rate that, at present, is likewise not
observable. This conclusion disagrees with earlier analyses. The
source of the disagreement is not in the treatment of the neutrino
evolution, but in the treatment of the solar magnetic fields, which we
believe were significantly overestimated in the earlier studies.

Overall, our conclusion is that, despite achieving a clearly
impressive level of sensitivity to $\bar\nu_e$ from the Sun, KamLAND
has not yet reached the level where it would be placing a meaningful
constraint on the neutrino transition moment and/or solar magnetic
fields. On the other hand, KamLAND has reached the point where it may
begin probing the ``optimistic'' scenario, $\mu_\nu\sim
10^{-11}\mu_B$. Clearly, further sensitivity improvements through
accumulation of additional statistics is very desirable.

\begin{acknowledgments}
  I owe special thanks to Andrei Gruzinov for countless -- always very
  clear and helpful -- discussions of the solar magnetic fields and
  plasma physics in general, to Matthias Rempel for a very helpful
  discussion and for pointing me to an excellent set of references,
  and to Cecilia Lunardini for many useful suggestions on the draft.
  I also acknowledge Stirling Colgate for a useful discussion
  of the physics of the convective zone fields. Additionally, I
  benefited greatly from stimulating correspondence with Evgeny
  Akhmedov and Timur Rashba that followed the initial posting of the
  preprint of this paper. This work was supported by the Department of
  Energy, under contract W-7405-ENG-36.
\end{acknowledgments}

\appendix

\section{Some technical details on the Convective Zone}

As described in Sect.~\ref{sect:cz}, simple physical arguments
give the following estimates for the probability of spin-flavor
flip in random magnetic fields:
\begin{eqnarray}
    P\sim\left\{
    \begin{array}{cc}
    (\mu B \lambda_{\rm corr})^2 L/\lambda_{\rm corr}, & \lambda_{\rm
    corr}\lesssim \lambda_{\rm osc}\\
   (\mu B \lambda_{\rm osc})^2 L/\lambda_{\rm corr}, & \lambda_{\rm
    corr}\gtrsim \lambda_{\rm osc}, \mbox{ sharp edge}\\
    \mbox{exp. suppressed,} & \lambda_{\rm
    corr}\gtrsim \lambda_{\rm osc}, \mbox{ smooth edge}
    \end{array}
    \right.
\end{eqnarray}
More generally, the probability of conversion is proportional to a
Fourier component of the field correlation function
\cite{rashbaprd}
\begin{equation}\label{eq:CZcorrfunc}
    P \simeq \mu^2 L \int_0^\infty dl' \langle B(0) B(l')\rangle \cos
    (2\Delta l').
\end{equation}
If $\langle B(0) B(l')\rangle$ is a slowly decaying function
(compared to $1/\Delta$) with a vanishing slope for small $l'$,
indicating no small-scale noise, for example $\langle B(0)
B(l')\rangle \propto \exp[-(l')^2/\lambda_{\rm corr}^2]$, the
resulting $P$ is exponentially small, as expected for adiabatic
evolution.

The quantity of interest in the isotropic turbulence model is
$B_{\lambda_{\rm osc}} \sim B_{L_{\rm max}}(\lambda_{\rm
osc}/L_{\rm max})^\alpha$. Refs. \cite{rashbaprl,rashbaprd}
performed an estimate based on a simple Kolmogorov turbulence.
Within this model, the effects of the magnetic field on the
character of the turbulence are neglected and the simple
Kolmogorov scaling law for velocities is applied, $v_\lambda \sim
(\epsilon \lambda)^{1/3}$, where $\epsilon$ is the energy
dissipation rate per unit mass. The field was assumed to scale
with $\lambda$ the same way as velocities, {\it i.e.},
$B_\lambda\propto \lambda^{1/3}$.

More generally, if the magnetic energy becomes comparable to the
kinetic energy of the fluid, the character of the turbulence changes
(see, e.g., \cite{zeldovich}, pp. 138-142). The rate of energy
transfer to small scales is decided by the interaction of Alfv\'en
waves in plasma. Recent investigations \cite{goldreich} argue that in
the presence of magnetic fields the turbulence becomes anisotropic in
$k$ space (see \cite{Gruzinov1997}, Appendix A, for a clear summary).
While a detailed discussion of the latest models of
magnetohydrodynamics turbulence is clearly beyond the scope of
this work, the preceding example show that the exponent in the scaling
law (\ref{eq:Bscaling}) is not known precisely. The correlation
function can then be parameterized as $\langle B(0) B(l')\rangle
\propto (l'/L_{\rm max})^\alpha K_\alpha(l'/L_{\rm max})$, yielding
upon integration
\begin{equation}
    P \simeq \mu^2 B_{L_{\rm max}}^2 (\lambda_{\rm osc}/L_{\rm max})^{2\alpha}
    L/\Delta,
\end{equation}
as expected on physical grounds.

\end{document}